\documentclass[12pt,journal,compsoc]{IEEEtran}
\usepackage{graphicx}
\usepackage{amssymb}
\usepackage{dcolumn}
\usepackage{bm}
\usepackage{psfrag}

\begin{document}

\title{Quantum information transfer in degenerate Raman regime} 

\author{T. Prud\^{e}ncio
\IEEEcompsocitemizethanks{\IEEEcompsocthanksitem Instituto de F\'\i sica, Universidade de Bras\'ilia - UnB, CP: 04455, 70919-970, Bras\'ilia - DF, Brazil.
and 
International Institute of Physics, Universidade Federal do
Rio Grande do Norte, Av. Odilon Gomes de Lima, 1722, 59078-400, Natal, RN, Brazil.\protect\\
E-mail: thprudencio@gmail.com}


\begin{abstract}
The interaction of a three level Rydberg atom of $\Lambda$-type with a single mode
optical field in far off-resonant and at large detuning regimes can be described by an effective degenerate Raman model, where the atomic state
can be treated as a two-level system of degenerate states. By means of this
approximation, we propose a quantum information transfer of an one-qubit
state from a Rydberg atom of $\Lambda$-type with degenerate levels to a single-mode
field initially in a coherent state. 
\end{abstract}

\markboth{submited to IEEE Journal of Quantum Electronics, October~2011}%
{Bare Demo of IEEEtran.cls for Computer Society Journals}

}

\maketitle
\IEEEdisplaynotcompsoctitleabstractindextext
\IEEEpeerreviewmaketitle

\section{Introduction}

Three-level atoms can appear in different configurations as $\Lambda$, $\Xi$ and V, what permits the realization of
population inversion with many applications in laser physics \cite{vedral, weisuter} and also lasing without inversion. By interacting with an one or two-mode 
optical field \cite{agarwal, gerry}, 
three-level atoms can in some special regimes be effectivelly described as two-level systems with adiabatic elimination of 
the highest level in the cases of $\Lambda$ \cite{wu} and $\Xi$ \cite{yang} configuration or the adiabatic elimination of
the lowest level in the case of V configuration \cite{xinhua}. 

For a $\Lambda$ configuration, 
a special case occurs when the two lower levels are degenerate. In the interaction with a single-mode optical field 
at far off-resonant and large detuning regimes the atomic states are reduced to 
a two-level system of degenerate states, characterizing a degenerate Raman regime. This effective interaction leads in fact to 
an adequate description of the Raman interaction in far off-resonant and large detuning regimes if compared 
to the full microscopic Hamiltonian of the Raman process \cite{xuluo}, in the case short evolving
times and description of physical quantities only involving the square of amplitude probabilities \cite{xu}.

From quantum optics, the degenerate Raman regime is well known, with important proposes of generation of non-classical states 
derived from it. For instance, the generation of Fock states \cite{avelarfilho, maiabaseia}, superpositions of coherent states \cite{zhengguo2}
 and superpositions of phase states \cite{avelarsouza}. In quantum information, protocols of quantum teleportation were proposed for unknown atomic states \cite{zhengguo}, 
unknown entangled coherent states \cite{feng} and superpositions of coherent states \cite{zheng}. 

In this paper, we propose a quantum information transfer of an one-qubit state from an atom to 
a single mode field initially in a coherent state. We consider an atom-field interaction 
in the degenerate Raman regime, such that the lower degenerated states of the atom are involved. Before the interaction, 
a measurement is realized in one of the atomic degenerate states and 
a Hadamard gate operation \cite{chuang} is realized into the single-mode field, realizing finally a quantum 
information transfer. This type of protocol was developed in other contexts \cite{cirac, feng2, zhang, paternostro, franco, parkins, li, wei} and it was
experimentally realized \cite{matsukevich, sherson}. In our case, the experimental realization can be achieved with three-level Rydberg atoms
 with principal numbers $49$, $50$ and $51$, adjusting the radiative times, atomic velocities 
and detuning to the single-mode coupling in the situation of degenerate Raman interaction 
\cite{zheng4, wang, hagley, raimond, osnaghi, davidovich}. On the other hand, Hadamard and NOT operations, one-qubit
 gate operations, can be implemented using linear-optical apparatus \cite{koshino}.

\begin{figure}[h]
\centering
\includegraphics[scale=0.4]{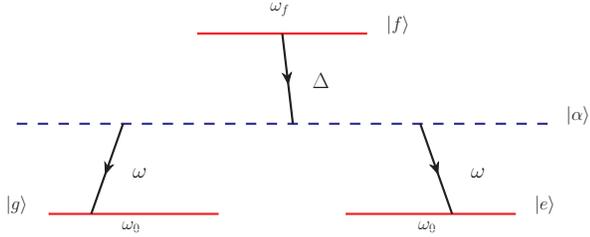}
\caption{(Color online) Scheme of degenerate Raman interaction.}
\label{gfr}
\end{figure}

We denote the three levels of the three-level Rydberg atom of $\Lambda$-type by
$|g\rangle$, $|e\rangle$ and $|f\rangle$. Due to the degeneracy, $|g\rangle$ and $|e\rangle$ have same frequency $\omega_{0}$. The frequency $\omega_{f}$ is associated to $|f\rangle$. The single-mode field is initially in a coherent state $|\alpha\rangle$ with frequency $\omega$ (See figure \ref{gfr}). Under atom-field degenerate Raman coupling the following relation is satisfied 
\begin{eqnarray}
\omega_{f}-\omega_{0} = \Delta + \omega,
\label{esg}
\end{eqnarray}
where $\Delta$ is the detuning between the atomic transition ( $\omega_{f} -\omega_{0}$) and the single-mode frequency $\omega$. 
In the case of large detuning, the upper state $|f\rangle$ can be adiabatically eliminated \cite{gerry}.

In the case of large detuning, short evolving times and physical quantities that only 
involve the square of the amplitude probabilities, the following effective Hamiltonian can be considered ($\hbar=1$) \cite{xuluo}
\begin{eqnarray}
\hat{H}_{ef}= \hat{n}\beta(|e\rangle \langle g|+\rm{h.~c.}),
\label{ef1}
\end{eqnarray}
where $\beta = -\lambda^{2}/\Delta$ is the effective atom-field coupling, $\lambda$ is the transition
 coupling from the lower states ( $|e\rangle$ and $|g\rangle$) to the upper state ($|f\rangle$), $\hat{n}=\hat{a}^{\dagger}\hat{a}$ 
is the number operator,
$\hat{a}$ and $\hat{a}^{\dagger}$ the creation and annihilation operators acting on the single-mode field.

In order to include effects of Stark shifts the term
\begin{eqnarray}
\hat{H}_{S}= \hat{n}\beta(|g\rangle \langle g|+|e\rangle \langle e|),
\end{eqnarray}
is added to the effective hamiltonian (\ref{ef1}), where we consider, for simplicity, the Stark parameters equal to the effective atom-field coupling $\beta$, such that we can write the effective degenerate Raman hamiltonian as \cite{xu}
\begin{eqnarray}
\hat{H} = \hat{H}_{S} + \hat{H}_{ef}.
\label{efs}
\end{eqnarray}
For the single-mode field initially in the coherent state $|\alpha\rangle$, the hamiltonian (\ref{efs}) has validity when the following inequalities \cite{xu} are satisfied
\begin{eqnarray}
\Delta^{2} \gg 2|2\lambda\alpha|^{2},
\label{d}
\end{eqnarray}
and
\begin{eqnarray}
t \ll \frac{3\Delta^{3}}{4|\lambda \alpha|^{4}}.
\label{b}
\end{eqnarray}

The time evolution during a time $t$ of an initial state of an atom in a superposed state of the form $c_{g}|g\rangle + c_{e}|e\rangle$,
$|c_{g}|^{2} + |c_{e}|^{2}=1$, and a field in a coherent state $|\alpha\rangle$ under the interaction (\ref{efs}) is given by 
\begin{eqnarray}
|\psi(t)\rangle &=& (c_{+}e^{-2i\hat{n}\beta t} -
c_{-})|g,\alpha\rangle \nonumber \\&+&(c_{+}e^{-2i\hat{n}\beta t} + c_{-})|e,\alpha\rangle,
\label{c}
\end{eqnarray}
where
\begin{eqnarray}
c_{\pm}=\frac{1}{2}\left(c_{e} \pm c_{g}\right).
\label{cpm}
\end{eqnarray}
We can also write the state (\ref{c}) as
\begin{eqnarray}
|\psi(t)\rangle &=& \left(c_{+}|e^{-2i\beta t}\alpha\rangle -c_{-}|\alpha\rangle \right)|g\rangle \nonumber \\
&+& \left(c_{+}|e^{-2i\beta t}\alpha\rangle +c_{-}|\alpha\rangle \right)|e\rangle.
\label{23e}
\end{eqnarray}
In the case $c_{g}=1$, $c_{e}=0$, corresponding to evolution of the ground state $|g\rangle$, we have
\begin{eqnarray}
|\psi(t)\rangle &=& \frac{1}{2}\left(|e^{-2i\beta t}\alpha\rangle +|\alpha\rangle \right)|g\rangle \nonumber\\
&+& \frac{1}{2}\left(|e^{-2i\beta t}\alpha\rangle -|\alpha\rangle \right)|e\rangle,
\end{eqnarray}
and in the case $c_{g}=0$, $c_{e}=1$, corresponding to evolution of the excited state $|e\rangle$, we have
\begin{eqnarray}
|\psi(t)\rangle &=& \frac{1}{2}\left(|e^{-2i\beta t}\alpha\rangle -|\alpha\rangle \right)|g\rangle \nonumber \\
&+& \frac{1}{2}\left(|e^{-2i\beta t}\alpha\rangle +|\alpha\rangle \right)|e\rangle.
\end{eqnarray}

\section{Quantum information transfer in degenerate Raman interaction regime}

The single-mode field is initially in a coherent state $|\alpha\rangle$. At degenerate Raman interaction regime, the upper level $|f\rangle$ of the three level Rydberg atom of $\Lambda$-type can be neglected, reducing to a two-level system described by the following state
\begin{eqnarray}
|\phi\rangle= c_{g}|g\rangle + c_{e}|e\rangle,
\label{atls}
\end{eqnarray}
where
\begin{eqnarray}
|c_{g}|^{2} + |c_{e}|^{2}=1.
\end{eqnarray}
In this way, the interaction between the single mode field $|\alpha\rangle$ and the atomic state $|\phi\rangle$ is given by the effective Raman interaction (\ref{efs}).

We want to realize the transfer of the unknown coefficients $c_{g}$ and $c_{e}$ from the atom to the single mode field in such a way that in the final step, we will have quantum state transfer from atom to the single mode field. As the atom in the form (\ref{atls}) corresponds to one qubit state, it will correspond to quantum state transfer of one qubit state.

The atom-field interaction occur during a time $t$, leading the system to achieve the following state
\begin{eqnarray}
|\psi_{\alpha\alpha'}\rangle &=& \left(c_{+}|\alpha'\rangle -c_{-}|\alpha\rangle \right)|g\rangle \nonumber \\
&+& \left(c_{+}|\alpha'\rangle +c_{-}|\alpha\rangle \right)|e\rangle,
\end{eqnarray}
where $c_{\pm}$ is given by (\ref{cpm}) and $\alpha'$ is related to $\alpha$ by means of
\begin{eqnarray}
\alpha'&=& e^{-2i\beta t}\alpha.
\label{time}
\end{eqnarray}
Measuring the atom in the excited state $|e\rangle$, the field is projected into the state
\begin{eqnarray}
|\phi_{+}\rangle = c_{+}|\alpha'\rangle +c_{-}|\alpha\rangle,
\end{eqnarray}
and the measurement of the atom in the ground state $|g\rangle$, the field is projected into the state
\begin{eqnarray}
|\phi_{-}\rangle = c_{+}|\alpha'\rangle - c_{-}|\alpha\rangle.
\end{eqnarray}
We choose the time of interaction $t=\pi/2\beta$ such that from (\ref{time}) we have
\begin{eqnarray}
|\phi_{\pm}\rangle = c_{+}|-\alpha\rangle \pm c_{-}|\alpha\rangle.
\label{finacod}
\end{eqnarray}
Now, taking into account the projection relations of coherent states \cite{walls}
\begin{eqnarray}
\langle \alpha | \alpha\rangle &=& 1,\\
\langle \alpha | -\alpha\rangle &=& e^{-2|\alpha|^{2}},
\end{eqnarray}
By chosing $|\alpha|$ suficiently large, but still satisfying degenerate Raman regime given by the inequalities (\ref{b}) and (\ref{d}) , we have
\begin{eqnarray}
\langle \alpha | -\alpha\rangle &\approx& 0.
\end{eqnarray}

In this case, $|\alpha\rangle$ and $|-\alpha\rangle$ are orthonormal states and the single mode field is described as a two-level system generated by the coherent states $|\alpha\rangle$ and $|-\alpha\rangle$ and capable of storing the qubit state expressed by the atomic state (\ref{atls}).

We note that the coeficients in (\ref{finacod}) do not correspond to $c_{g}$ and $c_{e}$, but are related to these by means of relations in (\ref{cpm}) or the following
\begin{eqnarray}
c_{+} + c_{-} &=& c_{e}, \\
c_{+} - c_{-} &=& c_{g}.
\end{eqnarray}
We then realize a Hadamard operation on the field state by means of the action the following operators $\hat{A}_{\alpha}$ defined by
\begin{eqnarray}
\hat{A}_{\alpha}&=&|-\alpha \rangle\langle -\alpha|-|\alpha \rangle\langle \alpha| 
+ |\alpha\rangle\langle -\alpha| \nonumber \\
&+& |-\alpha\rangle\langle \alpha|,
\end{eqnarray}
corresponding to a Hadamard operator \cite{chuang}
\begin{eqnarray}
\frac{1}{\sqrt{2}}\left(|0\rangle\langle 0| -|1\rangle\langle 1| + |1\rangle\langle 0| +|0\rangle\langle 1|\right),
\end{eqnarray}
where we disconsider the factors of $1/\sqrt{2}$ and make the correspondence $|0\rangle \rightarrow  |-\alpha\rangle$ and $|1\rangle \rightarrow |\alpha\rangle$.

\begin{figure}[h]
\centering
\includegraphics[scale=0.5]{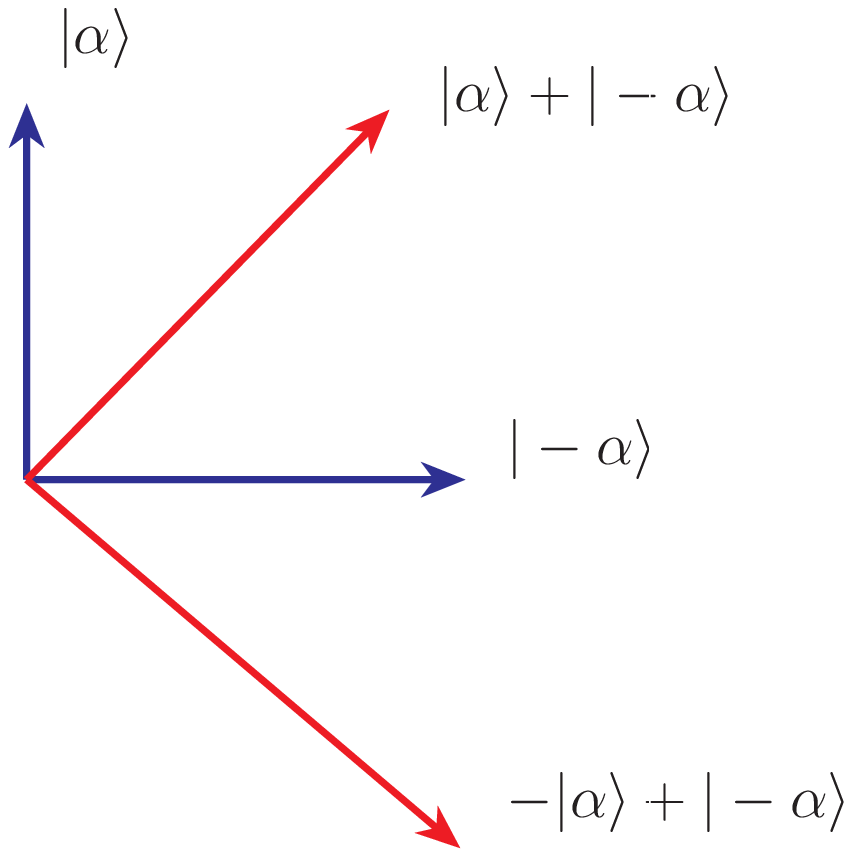}
\caption{(Color online) Scheme of $|\alpha\rangle$ and $|-\alpha\rangle$ states (blue arrows) and the transformations under Hadamard operation (red arrows).}
\label{gfrd}
\end{figure}

The operator $\hat{A}_{\alpha}$ acts on $|\alpha\rangle$ and $|-\alpha\rangle$ in the following way (see figure \ref{gfrd})
\begin{eqnarray}
\hat{A}_{\alpha}|\alpha\rangle= -|\alpha\rangle + |-\alpha\rangle,
\end{eqnarray}
\begin{eqnarray}
\hat{A}_{\alpha}|-\alpha\rangle= |\alpha\rangle + |-\alpha\rangle,
\end{eqnarray}

\begin{figure}[h]
\centering
\includegraphics[scale=0.5]{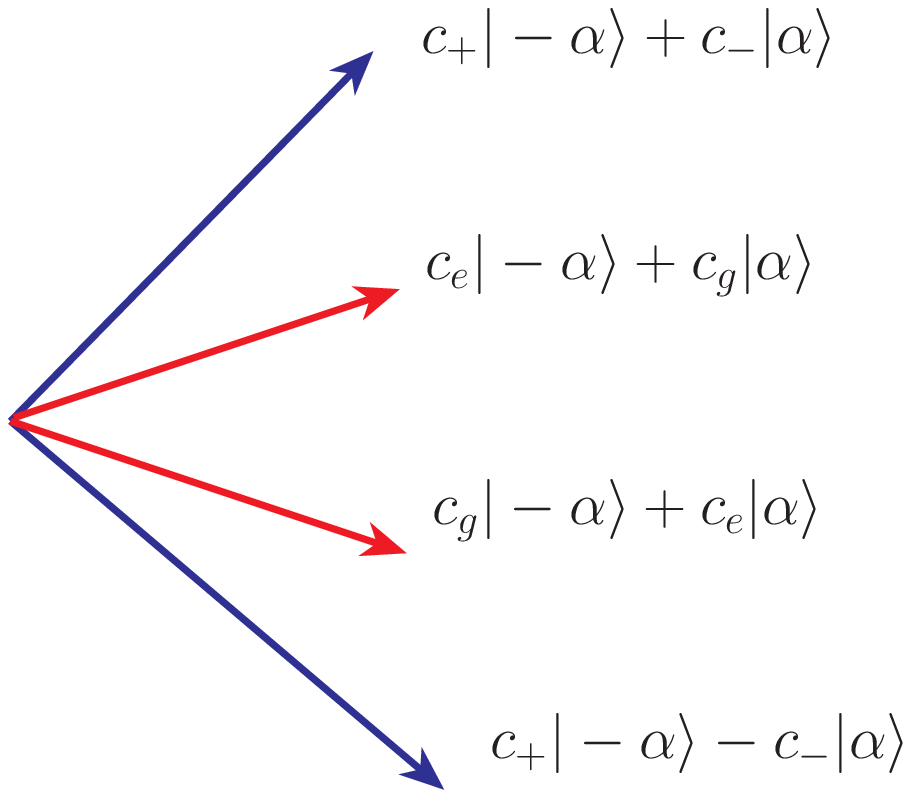}
\caption{(Color online) Scheme of $|\phi_{+}\rangle$ and $|\phi_{-}\rangle$ states (blue arrows) and the transformations under Hadamard operation (red arrows).}
\label{gfrd23}
\end{figure}

The action of the operator $\hat{A}_{\alpha}$ on the field $|\phi_{+}\rangle$ leads single mode field to the following state 
\begin{eqnarray}
\hat{A}_{\alpha}|\phi_{+}\rangle= c_{e}|-\alpha\rangle +c_{g}|\alpha\rangle,
\label{dasxz}
\end{eqnarray}
corresponding to a quantum information transfer when the atom is detected in the excited state $|e\rangle$.
On the other hand, the action of $\hat{A}_{\alpha}$ on the field $|\phi_{-}\rangle$ leads to the state 
\begin{eqnarray}
\hat{A}_{\alpha}|\phi_{-}\rangle = c_{g}|-\alpha\rangle +c_{e}|\alpha\rangle,
\label{dasxz2}
\end{eqnarray}
corresponding to a quantum information transfer when the atom is detected in the ground state $|g\rangle$ (see figure \ref{gfrd23}).

\section{Conclusion}

In conclusion, we have proposed a protocol of quantum information transfer where the atom-field interacts
in degenerate Raman regime, an atomic measurement in any one of the de-
generate states is realized and then a Hadamard gate operation is applied
into the field, storing in the single-mode field the qubit information present
initially in the atom.

Independent of the final state in which the atom is detected ( ground $|g\rangle$ or excited state $|e\rangle$), the action of the Hadamard operator $\hat{A}_{\alpha}$ on the field after the interaction leads to a quantum information transfer of the coefficients $c_{g}$ and $c_{e}$ from the atom to the single-mode field.

The order of the coefficients in (\ref{dasxz}) and (\ref{dasxz2}) does not 
matter in our protocol and their interchange could be realized by means of a NOT operation. 

Storage of quantum information and its carriage are fundamental problems 
to the realization of quantum computers, motivating important achievements. In this propose, a simple quantum information transfer
was proposed. This situation can be applied in quantum circuits where a qubit comes from 
an atom qubit state and is stored in a single-mode field qubit state.

\section*{Acknowledgments}
The author thanks CAPES (Brazilian government agency) for financial support.


\begin{thebibliography}{99}
\bibitem{vedral} V. Vedral, \textit{Modern Foundations of Quantum Optics} (Imperial College Press, London, 2005).
\bibitem{weisuter} C. Wei, D. Suter, A. S. M. Windsor, N. B. Manson, Phys. Rev. A, {\bf 58} (1998) 2310. 
\bibitem{agarwal} G. R. Agarwal, Phys. Rev. A, {\bf 1} (1970) 1445.
\bibitem{gerry} C.~C.~Gerry, J.~H.~Eberly, Phys.\ Rev.\ A {\bf 42} (1990) 6805. 
\bibitem{wu} Y. Wu, Phys.\ Rev.\ A {\bf 54} (1996) 1586.
\bibitem{yang} Y. Wu, X. Yang, Phys. Rev. A {\bf 56} (1997) 2443.
\bibitem{xinhua} Z. XinHua, Y. ZhiYong, X. PeiPei, Sci. China Ser. G Phys Mec. Atron. {\bf 52} (2009) 1034.
\bibitem{xuluo} L. Xu, Z-F. Luo, Z-M. Zhang, J. Phys. B, {\bf 27} (1994) 1649.
\bibitem{xu} L.~Xu, Z-M.~Zhang, Z.\ Phys.\ B {\bf 95} (1994) 507.
\bibitem{avelarfilho} A.~T.~Avelar, T.~M.~Rocha Filho, L.~Losano, B.~Baseia, Phys.\ Lett.\ A {\bf 340} (2005) 74.
\bibitem{maiabaseia} L. P. A. Maia, B. Baseia, A. T. Avelar, J. M. C. Malbouisson, J. Opt. B: Quantum. Semiclass. Opt. {\bf 6} (2004) 351.
\bibitem{zhengguo2} S-B.~Zheng, G-C.~Guo, Quantum Semiclass.~Opt.\ {\bf 9} (1997) L45.
\bibitem{avelarsouza} A. T. Avelar, L. A. de Souza, T. M. da Rocha Filho, B. Baseia, J. Opt. B: Quantum Semiclass. Opt. {\bf 6}  (2004) 383.
\bibitem{zhengguo} S-B.~Zheng, G-C.~Guo, Phys. Lett. A, {\bf 232} (1997) 171.
\bibitem{feng} M.~Feng, M.~A.~S.~She, Comm.\ Theor.\ Phys.\ {\bf 47} (2007) 330.
\bibitem{zheng} S-B.~Zheng, Chin. Phys. Journal, {\bf 42} (2004) 35.
\bibitem{chuang} M. A. Nielsen, I. L. Chuang, \textit{Quantum Computation and Quantum Information} (Cambridge University Press, Cambridge, 2000).
\bibitem{cirac} J. I. Cirac, P. Zoller, H. J. Kimble, H. Mabuchi, Phys. Rev. Lett., {\bf 78} (1997) 3221.
\bibitem{feng2} Z-B. Feng, Z-L. Cai, C. Zhang, L. Fan, T. Feng, Optics Communications, {\bf 283} (2010) 3221.
\bibitem{zhang} J. Zhang, K. Peng, S. L. Braunstein, Phys. Rev. A, {\bf 68} (2003) 013808.
\bibitem{paternostro} M. Partenostro, G. M. Palma, M. S. Kim, G. Falci, Phys. Rev. A, {\bf 71} (2005) 042311.
\bibitem{franco} C. Di Franco, M. Partenostro, M. S. Kim, Phys. Rev. A, {\bf 81} (2010) 022319.
\bibitem{parkins} A.S. Parkins, H. J. Kimble, J. of Opt. B: Quantum Semclass. {\bf 1} (1999) 496.
\bibitem{li} P-B. Li, Y. Gu, Q-H. Gong, G-C. Guo, Phys. Rev. A, {\bf 79} (2009) 042339.
\bibitem{wei} L. F. Wei, J. R. Johansson, L. X. Cen, S. Ashhab, F. Nori, Phys. Rev. Lett., {\bf 100} (2008) 113601.
\bibitem{matsukevich} D. N. Matsukevich, A. Kuzmich, Science {\bf 85} (2004) 306.
\bibitem{sherson} J. F. Sherson, H. Krauter, R. K. Olsson, B. Jusgaard, K. Hammerer, I. Cirac, E. S. Polzik, Nature, {\bf444} (2006) 557.
\bibitem{zheng4} S-B.Zheng, G-C. Guo, Phys. Rev. Lett. {\bf 85} (2000) 2392.
\bibitem{wang} X-W.~Wang, G-J.~Yang, Optics Communications {\bf 281} (2008) 5282.
\bibitem{hagley} E. Hagley, et. al., Phys. Rev. Lett., {\bf 79} (1997) 1.
\bibitem{raimond} J. M. Raimond, M. Brune, S. Haroche, Rev. Mod. Phys., {\bf 73} (2001) 565.
\bibitem{osnaghi} S. Osnaghi, P. Bertet, A. Auffeves, P. Maioli, M. Brune, J. M. Raimond, S. Haroche, Phys. Rev. Lett. {\bf 87} (2001) 037902.
\bibitem{davidovich} L. Davidovich, A. Maali, M. Brune, J. M. Raimond, S. Haroche, Phys. Rev. Lett. {\bf 71} (1993) 2360.
\bibitem{koshino} K. Koshino, S. Ishizaka, Y. Nakamura, Phys. Rev. A, {\bf 82} (2010) 010301.
\bibitem{walls} D. F. Walls, G. J. Milburn, \textit{Quantum Optics} (Springer-Verlag, Berlin, 2008)
\end{thebibliography}
\end{document}